\documentstyle[aps,prl,epsfig,twocolumn]{revtex}

\begin{document}
\draft

\twocolumn[\hsize\textwidth\columnwidth\hsize\csname @twocolumnfalse\endcsname

\title{Microscopic theory of vibronic dynamics in linear polyenes}
\author{L. Arrachea$^{1,*}$, A.A. Aligia$^{2}$, and G.E. Santoro$^{1}$}

\address{
$^{1,*}$ Scuola Internazionale Superiore di Studi Avanzati (SISSA) and
Istituto Nazionale per la Fisica della Materia (INFM) (Unit\`a di Ricerca
Trieste-SISSA), Via Beirut 4, I-34014 Trieste, Italy\\
$^2$ Centro At\'omico Bariloche, (8400) Bariloche, Argentina.}

\date{\today}
\maketitle
\begin{abstract}

We propose a novel approach to calculate dynamical processes at ultrafast 
time scale in molecules in which vibrational and electronic motions are 
strongly mixed. The relevant electronic orbitals and their interactions
are described by a Hubbard model, while electron-phonon interaction terms
account for the bond length dependence of the hopping and 
the change in ionic radii with valence charge. 
The latter term plays a crucial role in the  non-adiabatic internal 
conversion process of the molecule.
The time resolved photoelectron spectra are in good 
qualitative agreement with experiments.

\end{abstract}

]


Photoexcitation in polyatomic molecules leads to the rapid mixing 
of vibrational and electronic motions, inducing charge redistribution 
and energy flow in the molecule. This non-adiabatic coupling is
essential in photochemical processes\cite{mich}, in photobiological 
processes, such as those involved in vision\cite{scho}, and in 
molecular electronics\cite{jort}. 

The progress in the technology of laser pulses has triggered 
the development of a new generation of 
spectroscopies devoted to the investigation of ultrafast phenomena.
Beautiful experiments have been recently
reported on the study of ultrafast non-adiabatic processes in
isolated molecules\cite{expe,nat}. 
Time-resolved photoelectron spectroscopy (TRPS) has been used 
to follow the vibronic dynamics of  all-{\em trans} 2,4,6,8 decatetraene 
\cite{nat}.   
In this molecule, the electronic ground state is the singlet S$_0$ 
($1^1$ A$_g$), 
while the first optically allowed excited state, S$_2$ ($1^1$ B$_u$), 
has an energy higher than the first excited ``dark'' singlet S$_1$ 
($2^1$ A$_g$). 
A laser pulse (pump) prepares the molecule in a vibrationally hot
wave packet involving the state S$_2$. 
The packet evolves and its time evolution is 
probed by photoexciting an electron and analyzing the ensuing spectra 
at subsequent times.
The population of 
the ``dark'' band, associated with the S$_1$ electronic state increases
with time, as a consequence of a non-adiabatic internal conversion between 
vibrational and electronic excitations. 
Such a process lasts a few hundred femptoseconds, and is a manifestation 
of the failure of the adiabatic description.

The aim of any theory in this field is to relate experimental 
spectroscopical data 
with the basic microscopic interactions of the system. 
This task is particularly difficult in the present case.
A theory describing the experiments should start from a model which
includes both electron-electron correlations as well as electron-phonon 
couplings, and should treat the latter beyond the adiabatic approximation. 
So far, to the best of our knowledge,
time dependent photoemission spectra for such non-adiabatic 
internal conversion processes have never been calculated on the
basis of a miscroscopic model. Previous
theoretical work on the internal conversion process is based 
on semiempirical models for the relevant energy surfaces, supplemented
by phenomenological couplings between them\cite{ger}.

The goal of this Letter is to describe the whole process behind a TRPS 
experiment \cite{nat} (laser pump followed by a dynamical internal conversion,
probed by the ensuing time-dependent photoemission spectra), 
using a minimal microscopic model in which both electron-electron and 
electron-phonon interactions are exactly taken into account.
Our results show a good qualitative agreement with the experimental 
findings \cite{nat},  and
provide a link between these and the basic underlying microscopic
interactions.

The model we study is based on a formal separation of the hybridized s and p 
orbitals lying in the plane of the molecule 
(see Fig. 1(a)), 
and the p-$\pi$ orbitals
perpendicular to it. 
The former are those with the larger
contribution to the chemical bond and vibration dynamics. The latter 
are those involved in the low-energy excitations of interest to us, and
are described 
by a Hubbard Hamiltonian. Its key ingredient is a strong local Coulomb 
interaction $U$.
Electron-phonon coupling terms are included, as derived from 
an expansion of the electron-ion interaction up to first order in the ionic 
displacements relative to the equilibrium positions \cite{vibl}.
The model reads:
\begin{eqnarray}
{\cal H} &=& -\sum_{\sigma=\uparrow,\downarrow}
\sum_{i=1}^{L-1}[t_0- g(u_{i+1}-u_i)]
(f^{\dagger}_{i,\sigma}f_{i+1,\sigma}+H.c) \nonumber\\
& & +U \sum_{i=1}^{L} n_{i\uparrow} n_{i\downarrow} 
+g^{\prime}\sum_{i=1}^{L-1} (u_{i+1}-u_i)(q_i+q_{i+1})
\nonumber\\
& &+ \frac{K}{2}\sum_{i=1}^{L-1} \sum(u_i-u_{i+1})^2 + 
\frac{1}{2m}\sum_{i=1}^L P_i^2 \;,
\label{e1}
\end{eqnarray}
where $i=1,\ldots,L$ labels the $\pi$ orbitals of the C atoms,
$f^{\dagger}_{i,\sigma}$ creates an electron with spin $\sigma$ at  
site $i$, 
$n_{i\sigma}=f^{\dagger}_{i,\sigma}f_{i,\sigma}$ is the corresponding 
number operator, 
and $q_i=1-n_{i\uparrow}-n_{i\downarrow}$ is the net charge at site $i$.
$u_i$ are the ionic displacements with respect to the equilibrium 
positions, and, for simplicity, the vibration dynamics 
has been assumed to be that of a one dimensional chain of ions (the CH units)
of mass $m$ with a nearest-neighbor harmonic constant $K$. 
Two electron-phonon interaction terms have been included in Eq.\ (\ref{e1}),
with coupling constants labeled by $g$ and $g^{\prime}$. 
The $g$-term is quite standard and accounts for the fact that 
the magnitude of electronic nearest-neighbor hopping becomes weaker 
or stronger as the bond stretches or compresses, respectively. 
The $g^{\prime}$-term, with $g^{\prime}>0$, accounts for the contraction 
(expansion) of the bond upon removal (addition) of an electron. 
Such a term is a consequence of the dependence upon the valence charge 
of the spread of the wave functions (particularly the $\sigma$ orbitals), 
a spread which in turn modifies the bond lengths.

For sake of simplicity, and in order to be able to numerically solve 
{\em exactly} the Hamiltonian, we consider a molecule with $L=4$ C 
atoms. 
The vibrational motion of the molecule 
for $g=g^{\prime}=0$
can be described in terms
of three normal modes, $e_a$, $e_b$, and $e_c$, oscillating with
frequencies $\omega_a=\sqrt{(2+\sqrt{2})K/m}$,
$\omega_b=\sqrt{(2-\sqrt{2})K/m}$, and 
$\omega_c=\sqrt{2K/m}$, as sketched in Fig. 1(b).
The mode $e_b$ will not be included, since it describes a 
uniform dilatation or contraction of the molecule, which should have a small 
coupling with the electrons in a larger molecule.

In order to show that the Hamiltonian (\ref{e1}) represents
the minimal model containing the relevant ingredients to describe the
physics of decatetraene, we begin by discussing the role played by each 
interaction term within the adiabatic picture.

{\em Role of $U$.}
The role of electronic correlations in the  excitation 
spectrum of polyenes has been already discussed \cite{bar,ross}. 
Here if $U$ were neglected,  the 
optically allowed $S_2$  state would have a lower energy than 
the dark $S_1$ state,
while  
$(U/t_0)^{>}_{\sim}1.7$ leads to the observed ordering 
of both singlet excitations.
Given the reasonable estimate of $t_0=2$ eV for the electronic 
hopping\cite{war}, we find that the value of $U$ that best fits the
data on decatetraene is $U=4.2$ eV.

{\em Role of $g$.}
The electron-phonon interaction $g$ is a standard ingredient of the
Pariser-Parr-Pople-Pierls (PPPP) model\cite{bar,ross}, crucial in 
explaining dimerization effects.  
The coupling between the electrons and the $e_a$ 
mode (the dimerization mode) leads to different equilibrium lengths 
for C-C and C=C bonds, alternating short and long bonds. 
From experimental data \cite{zer} we took $\omega_a=0.2$ eV 
(which implies $K=36$ eV/$\AA^2$ and $\omega_c=0.15$ eV).
A value for $g=3$ eV/$\AA$ is then calculated by fitting the 
experimentally observed difference in bond-lengths\cite{zer}.
Setting for the time being
$g^{\prime}=0$, one can study the 
Born-Oppenheimer (BO) surfaces 
corresponding to the three lowest energy singlets S$_0$, S$_1$, S$_2$, 
as functions of the normal mode coordinates $e_a$ and $e_c$ (see Fig.\ 1(c)).
The position of the minimum of the ground state BO surface,
occurring at a finite $e_a>0$, reflects the dimerization. 
It is also evident from Fig.\ 1(c) that the electron-phonon coupling 
introduces anharmonicities in the BO energy surfaces leading to the 
occurrence of level crossings in the excited states, known as
conical intersections\cite{vibl}.
The S$_2$ singlet is odd under space inversion while the S$_1$ is even. 
The conical intersection between both states is a central feature of the 
physics we want to describe.
Note that the $e_a$ mode is even under inversion, while $e_c$ is odd. 
Then, on general symmetry grounds, the coupling between the electrons 
and the $e_c$ mode would be expected to produce a nonvanishing matrix element 
coupling the S$_1$ and the S$_2$ states, leading eventually to an
{\em avoided crossing} between the corresponding bands, with
a gap between the two proportional to the effective interband 
tunneling amplitude\cite{vibl}. 
As a consequence of that mixing, the adiabatic description in terms of 
uncoupled BO surfaces would loose meaning, and the quantum nature of 
the phonons should be included explicitly.
However, in absence of the $g^{\prime}$-term, the Hamiltonian possesses
a subtle electronic symmetry (particle-hole $P_{p-h}$) 
which leads to the vanishing of the 
interband tunneling matrix element, even for $e_c\ne 0$, as explained below.
$P_{p-h}$ is defined 
as the invariance of ${\cal H}$ 
(when $g^{\prime}=0$, and up to an inessential chemical potential shift) 
under the transformation 
$f^{\dagger}_{i,\sigma} \rightarrow (-1)^i f_{i,-\sigma}$. 
Taking into account that the electronic states in polyenes correspond 
to half-filled configurations (the number of $\pi$-electrons $N=L$)
and that this subspace is left invariant by $P_{p-h}$,
the eigenstates of ${\cal H}$ within this sector can be
classified as even or odd according to $P_{p-h}$ \cite{bar}. 
Detailed analysis reveals that S$_1$ is even under $P_{p-h}$,
while S$_2$ is odd. 
As a consequence, the interaction term $g$ alone cannot produce the 
interband coupling leading to the non-adiabatic effects observed in 
these molecules.

{\em Role of $g^{\prime}$.}
For this reason, the additional interaction $g^{\prime}$, which breaks 
particle-hole symmetry, is {\em essential} for the internal conversion
in our model. 
Fig.\ 1(d) shows the BO energy surfaces when $g^{\prime}$ is included. 
Notice that the ground state energy surface is practically unaffected 
by this interaction. This is due to the fact that the on-site Coulomb 
repulsion inhibits charge fluctuations in the ground state, thus
making the $g^{\prime}$ term, involving the net charges $q_i$, ineffective.
On the contrary, charge fluctuations are important in the excited states, 
and couple to the phonon modes through the $g^{\prime}$ term: 
As a consequence of the quantum
tunneling for $e_c\neq 0$, due to $g^{\prime}$, the two relevant 
potential-energy surfaces show now an avoided crossing (Fig.\ 1(d)), 
instead of intersecting (Fig.\ 1(c)).
A rough estimate for the strength of $g^{\prime}$ can be obtained 
from its effect on a C$_2$ dimer. 
By fitting experimental data for the bond lengths 
of C$_2$ (1.2425 $\AA$) and of C$_2^-$ (1.2682 $\AA$) with our model, we 
get $g^{\prime}\approx 3.95$ eV$/\AA$ \cite{tabla}.
An alternative rough estimate based on the ionic radii of 
C$^{+4}$ and C$^{-4}$ ions (0.15 and 2.60 $\AA$ respectively \cite{sp}), 
leads to $g^{\prime}\approx 10$ eV$/\AA$. We found that 
$g^{\prime}\approx 5$ eV/$\AA$ leads to reasonable values for  
the effective interband coupling and for the position of the conical 
intersection.

In order to describe the observed internal conversion 
effects, the adiabatic picture must be abandoned. The normal coordinates 
are quantized as:
%
$e_a=\sqrt{{\hbar}/{2m\omega_a}}(a+a^{\dagger}), \;
e_c=\sqrt{{\hbar}/{2m\omega_c}}(c+c^{\dagger})$.
%
The resulting Hamiltonian ${\cal H}$ can be numerically diagonalized by 
introducing a cutoff in the number of phonons $n_a$ and $n_c$. 
We found that keeping states with up to $n_a=n_c=10$ was enough to obtain 
accurate results.

The laser pump is simulated by acting at $t=0$ on the ground
state $|\phi_0\rangle$ with the following operator:
\begin{equation}
\hat{O}= \sum_{m} G_m |\phi_m\rangle \langle \phi_m| \hat{E},
\label{e3}
\end{equation}
where $|\phi_m\rangle$ denote the exact eigenstates of ${\cal H}$, 
and the effect of the laser electric field is represented by the dipole 
operator 
$\hat{E}=\sum_i R_i q_i$, where $R_i$ is the position of the C atom $i$ for
$u_i$=0.
The pump pulse is assumed to have a Gaussian shape in time. 
The Fourier transform of its envelope determines the excitation amplitudes 
$G_m \sim \exp{(-(\epsilon_m-\overline{\epsilon})^2/2\sigma_p^2)}$, 
where $\epsilon_m=E_m-E_0$ are the excitation energies relative to the ground
state $E_0$, while $\overline{\epsilon}$ and $\sigma_p$ are
the mean excitation energy and the energy spread of the 
pump pulse. 
The Schr\"odinger time evolution of the prepared wave packet is then given by:
%
$|\psi(t)\rangle=\exp{({i} {\cal H} t/{\hbar}}) \hat{O} |\phi_0\rangle.$
%
A qualitative picture of the resulting dynamics 
can be obtained from 
$P_o(t)=\langle \psi(t)|P_o|\psi(t)\rangle$, where $P_o$ is defined
as the projector of the eigenstates of ${\cal H}$ on the subspace of electronic
states with odd parity under space inversion. 
$P_o(t)$  is shown in Fig.\ 2. 
The first two singlet excitation energies 
(with predominantly S$_1$ and S$_2$ character) are 
$\sim 3.74$ eV and $\sim 4.31$ eV respectively. 
The spectral density profile of the initial wave packet 
is shown in the inset of Fig.\ 2 for a pulse with $\sigma_p=2$ eV and 
$\overline{\epsilon}=4.3$ eV. 
The results for $P_o(t)$ in Fig.\ 2 show that, on top of a
fast oscillatory component, there is a slower internal conversion component
which lasts approximately $243 fs$. 
This process is dominated by the two 
excited states labelled by (1) and (2) in the inset of Fig.\ 2. 
The excitation energies of these states are $\epsilon_1=4.304$ eV and 
$\epsilon_2=4.315$ eV, respectively. 
State (1) has predominantly S$_1$ (even) character 
($\langle\phi_1|P_o|\phi_1\rangle \sim 0.4$), while
state (2) has predominantly S$_2$ (odd) character 
($\langle\phi_2|P_o|\phi_2\rangle \sim 0.74$). 
The evolution of a wave packet composed by just these two states is indicated 
with a dashed line in Fig.\  2. 
This behavior suggests that the relevant time scale in the evolution of 
$|\psi(t)\rangle$ is set by $\sim \hbar/(\epsilon_2-\epsilon_1)$.

Within the sudden approximation, the photoelectron spectrum at time $t$
is essentially a measure of the following spectral function
\cite{fetter}
\begin{equation}
\rho_t(\omega) = \sum_{i,\sigma} 
\sum_{m}|\langle \phi^{\prime}_m|f_{i,\sigma}|\psi(t)\rangle|^2
\delta(\omega-E^{\prime}_m),
\label{e5}
\end{equation} 
where $E^{\prime}_m$ and $ |\phi^{\prime}_m\rangle$ denote the 
eigenenergies and eigenstates of ${\cal H}$ upon removal of a valence 
electron.   
Results for the time evolution of $\rho_t(\epsilon)$ are shown in
Fig.\  3, where an artificial broadening of the delta-functions has been 
introduced.
The most salient feature of Fig.\ 3 is the transfer of spectral weight, 
as a function of time, from a group of states close in energy to the ground 
state of the system with $N=L-1$ electrons 
(all of them with a predominantly D$_0$ nature, {\it i.e.}\ a doublet obtained 
by photoexciting $S_2$ \cite{nat}), 
to a group of states at a higher energy.
The latter states are identified as vibrationally hot states with predominantly
D$_1$ nature (obtained by photoexciting $S_1$).
The arrow in the lowest panel of Fig.\ 3 indicates the position of the 
lowest-energy state related to such a band.  
The behavior of the photoelectron spectra as a function of time 
is in qualitative agreement with the experimental results of TRPS. 
The larger energy gap between the relevant photoelectron spectral features
which we find in our simulation 
($\sim 2$ eV in our case, to be compared with the experimental result
of $\sim 1.2$ eV \cite{nat}) 
should be likely ascribed to the smaller size of the molecule
we are considering in our simulation.    

In summary, we have shown that the model Hamiltonian in Eq.\ (\ref{e1}) 
contains all the basic ingredients which are necessary to explain the 
essential features of the the dynamics of linear polyenes, notably
electron-electron and electron-phonon interactions treated in a 
non-adiabatic (fully quantum) framework. 
With such a model, we have described in detail the various features of a 
time-resolved photoemission spectroscopy experiment, 
obtaining a good picture of the underlying physics. 
We remark that, although the ground state properties as well as the 
structure of the electronic excited states are properly described by 
the standard PPPP model \cite{bar,ross}, 
the proper treatment of the 
observed non-adiabatic internal conversion, related to tunneling between 
coupled Born-Oppenheimer states, requires the inclusion of usually neglected 
electron-phonon interactions, such as the $g^{\prime}$ term we have considered,
which break a residual particle-hole symmetry of the Hamiltonian.
On the technical side, we remark that exact diagonalization techniques are   
useful theoretical tools for the study of these effects, as shown by the
reasonable agreement of our simulations with the behavior reported in Ref.\ 
\cite{nat}.

We thank P. Bolcatto for useful discussions. LA and AAA acknowledge support 
from CONICET. Part of the numerical work was done at the Max Planck 
Institut PKS. GES acknowledges support by MIUR under project COFIN.

\begin{figure}
\epsfxsize=3.2in
\epsffile{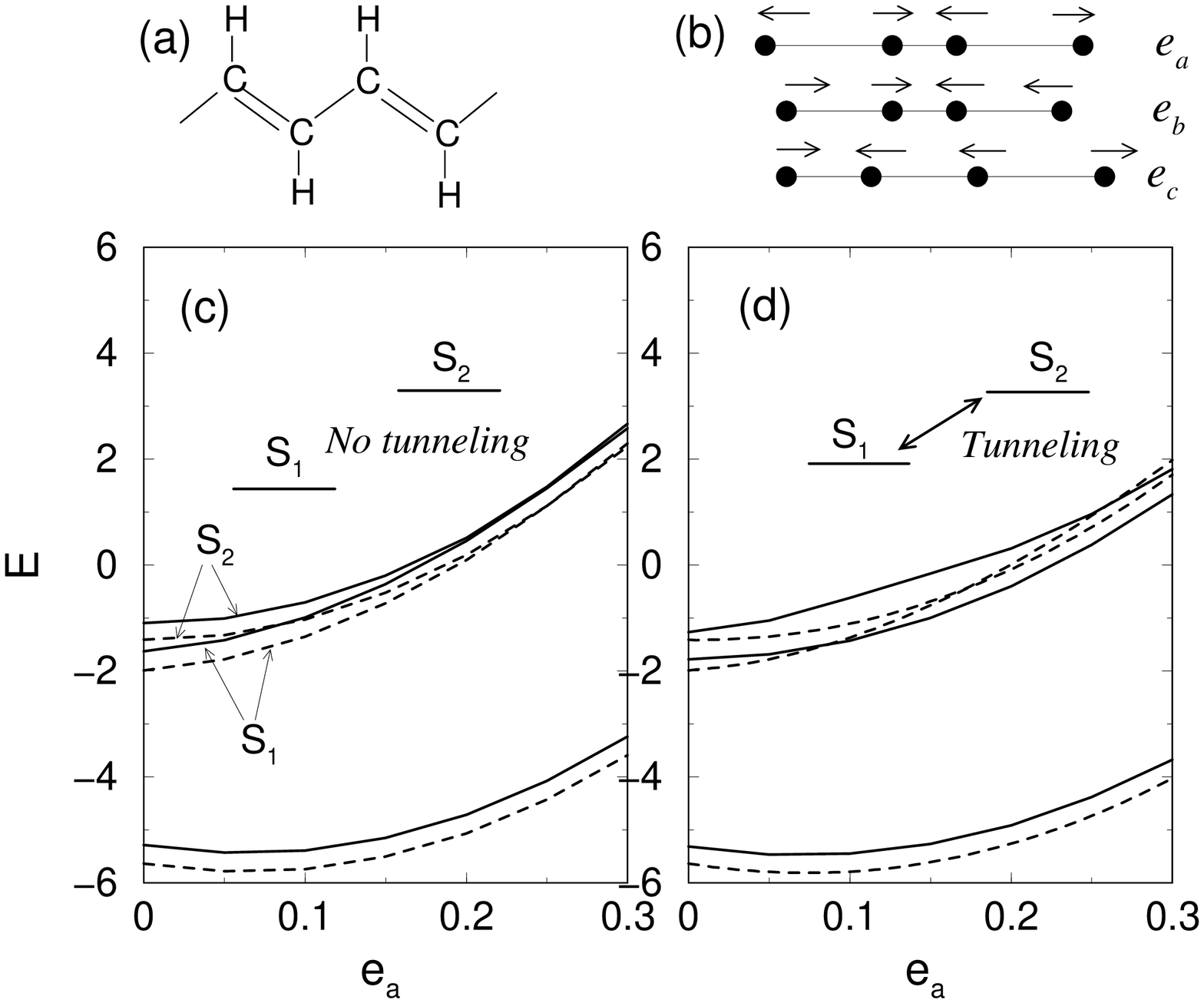}

\caption{
(a) Scheme of the linear polyene. Carbon and Hydrogen atoms  are
denoted by C and H respectively. (b)
Scheme of the normal modes of the molecule. (c) Cuts of the potential
energy surfaces for $g^{\prime}=0$, with
$t_0=2$ eV, $U=4.2$ eV, $g=3$ eV/$\AA$, $K=36$ eV/$\AA^2$. 
Dashed and solid lines correspond
to $e_c=0,0.1$, respectively. (c) Same as (b) for $g^{\prime}=5$ eV/$\AA$. }
\label{fig1}
\end{figure}

\begin{figure}
\epsfxsize=3.2in
\epsffile{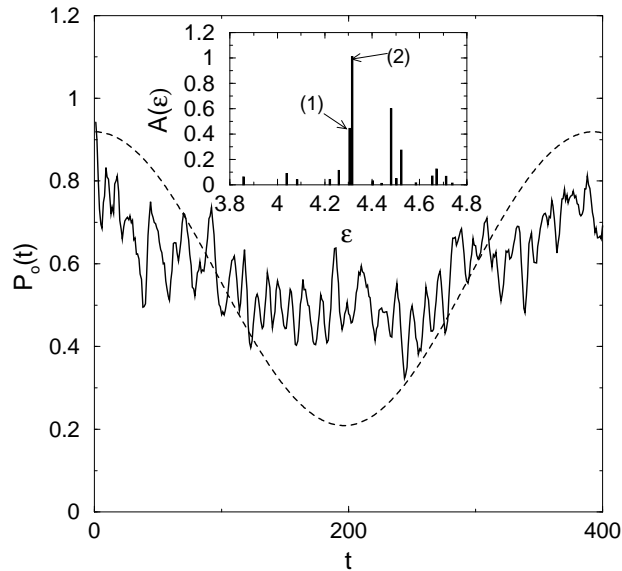}

\caption{Evolution of the projector $P_o(t)$ (Solid line).
The model parameters are those of Fig. 1(c).
Dashed line: evolution of $P_o(t)$ for a wave packet consisting of the
two states (1) and (2) indicated in the inset.
Inset: Spectrum of the prepared wave packet for a Gaussian
laser pump pulse with $\overline{\epsilon}=4.3$ eV and $\sigma_p=2$ eV.}
\label{fig2}
\end{figure}

\begin{figure}
\epsfxsize=3.2in
\epsffile{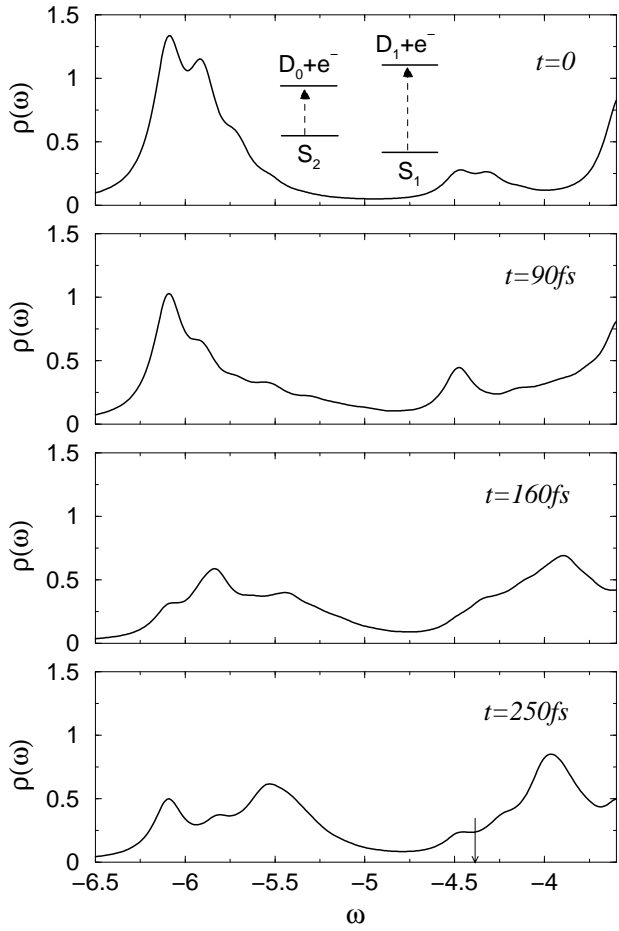}
\caption{Evolution of the photoelectron spectra for the
prepared wave packet of Fig.  2.}
\label{fig3}
\end{figure}

\end{document}